\documentstyle[eqsecnum,preprint,aps]{revtex}

\begin{document}
\draft
\title{Progetto per la costituzione del\\
CENTRO DI DIVULGAZIONE DELLA\\
CULTURA SCIENTIFICA\thanks{%
Su questo progetto il Comune di Sesto Fiorentino (Provincia di Firenze,
Italia) sta basando gran parte del suo intervento negli Istituti scolastici
pubblici del proprio territorio, nel quadro della Riforma dell'Autonomia
Scolastica varata dal ministro L. Berlinguer dei Governi Prodi e D'Alema.}\\
di Sesto Fiorentino\thanks{%
The English version is going to be published soon as possible!}.}
\author{M. Bertini, F. Gelli, M. Materassi, L. Mordini,}
\address{R. Nibbi, G. Paoli, M. Torricini, M. Ulivi\thanks{%
The authors Bertini, Gelli, Materassi, Nibbi, Paoli and Torricini are
members of the association referred to as {\it S.E.T.} (Scienza Educazione
Territorio), while the authors Mordini and Ulivi are members of the
association referred to as {\it Galea} (Navigare nella Cultura).}}
\address{materassi@fi.infn.it}
\date{\today }
\maketitle

\begin{abstract}
In questo articolo due piccole associazioni culturali di giovani scienziati
dell'hinterland fiorentino espongono il loro progetto per la costruzione di
un Ente culturale scientifico a carattere locale, fortemente ancorato alle
caratteristiche specifiche della loro Citt\`a, Sesto Fiorentino.

Ne esce il disegno di una iniziativa di privato sociale, il Centro per la
Divulgazione della Cultura Scientifica, operante in forte sinergia con
l'Amministrazione cittadina e le Scuole pubbliche del territorio.
\end{abstract}

\newpage\ 

\section{Premessa.}

Una delle caratteristiche pi\`u peculiari ed importanti della ''societ\`a
civile'' nella Citt\`a di Sesto Fiorentino \`e senza dubbio la presenza
massiccia di associazioni; colpisce non soltanto il loro numero,
letteralmente sterminato in rapporto agli abitanti, ma anche la loro
variet\`a per carattere organizzativo, per interesse, per obbiettivi.

Colpisce il rilievo storico delle Case del Popolo sestesi, dei gruppi
parrocchiali, le iniziative ricreative ed umanitarie a cui danno vita; la
tradizione della Venerabile Misericordia. E colpisce soprattutto la
vitalit\`a con cui, continuamente, nascono e crescono nuove associazioni,
sportive e culturali e politiche, che fioriscono vivissime a fianco dei
grandi protagonisti che le hanno precedute.

Per questo Sesto Fiorentino dev'essere considerato un esempio unico di
Citt\`a in cui \`e facile ed incoraggiante costruire iniziative nuove in
campo associativo, in cui una tradizione lunga, ed una lungimirante
Amministrazione Comunale, hanno permesso l'evolversi di questa grande
ricchezza.

In questa Citt\`a esistono molte proposte di iniziativa culturale
nell'ambito umanistico: la tradizione italiana del resto vede l'evoluzione
culturale di cittadini e lavoratori avvenire soprattutto attraverso la
letteratura, la pittura, il teatro, la musica, le forme di comunicazione che
narrano di quella stessa gente, delle sue lotte, delle sue condizioni, dei
suoi sogni. Questo panorama ultimamente si \`e sviluppato in una nuova
direzione, rispondendo a trasformazioni culturali di tutta la societ\`a.

Le societ\`a industriali si trasformano sempre pi\`u rapidamente, al passo
delle nuove tecnologie, e questo produce nella cultura due effetti collegati
e solo apparentemente antitetici: l'interesse per le scienze esatte, per gli
aspetti anche pi\`u difficili di tali discipline (di cui le tecnologie e i
progressi sono sottoprodotto), il richiamo del mondo naturale, la nascita
dell'attenzione agli equilibri ecologici (che con quei progressi spesso sono
entrati in drammatico conflitto). C'\`e un risveglio, una curiosit\`a nuova
verso gli esseri viventi vegetali ed animali, c'\`e una coscienza ecologica
che si fa faticosamente strada, che lotta per venir fuori, per chiedere alla
cultura spiegazioni ed atteggiamenti nuovi, valori diversi; c'\`e
un'attenzione alla propria salute, ed ai propri diritti in materia
ambientale, che fa il cittadino pi\`u curioso di come \`e governato il
territorio in quanto tale, e la curiosit\`a diviene richiesta, istanza,
mobilitazione. Il mondo associativo registra questo risveglio, si sono
costituite associazioni nuove rivolte all'ambiente, alla cultura
naturalistica.

Pi\`u difficile, perch\'e meno ''ludico'' forse, \`e il compito di gruppi
culturali che rivolgono la loro attenzione alla diffusione delle scienze
esatte di per s\'e; nel 1990 prende avvio a Sesto Fiorentino l'attivit\`a di
Math90, operante attorno alla Biblioteca pubblica, con l'ambiziosa
intenzione di fare della matematica materia di socializzazione,
divertimento, approfondimento culturale ''ricreativo''; nel 1998 due giovani
associazioni, Galea e Perseverare Ovest, operanti a livello intercomunale
fra Sesto, Campi e Calenzano, danno vita ad una mostra in cui gli studenti
delle scuole dell'obbligo e superiori entrano in contatto con fatti
elementari e non della Fisica e della Chimica, suscitando un notevole
interesse, spia delle grandi possibilit\`a che offre, a Sesto Fiorentino, il
campo delle divulgazione, ''ricreativa'' e non solo, delle scienze.

Di lavoro nell'ambito della diffusione della cultura scientifica, come si
vede, ce n'\`e da fare: la societ\`a in cui viviamo richiede un numero
sempre maggiore di tecnici specializzati nelle pi\`u varie discipline, ma la
troppa specializzazione porta spesso a perdere il contatto con la scienza
nel suo insieme, e chi sa tutto su un argomento spesso ignora tutto il
resto. Tale effetto porta alla formazione di coltissime trib\`u tecnologiche
chiuse in se stesse, talora diffidenti o concorrenti le une delle altre, e
questo \`e un regresso, e non un progresso, del vivere civile: questa
scienza, patrimonio di professionisti settari, \`e vista dai non adepti come
qualcosa di estremamente misterioso e oscuro, tanto da rappresentare per
molti ci\`o che nel medioevo erano alchimia e magia. Questo pu\`o essere
combattuto cercando di avvicinare la popolazione alla scienza in maniera
semplice e intuitiva, collegando fenomeni osservabili quotidianamente alle
leggi della Fisica, della Chimica e della Matematica.

Nonostante il problema dell'inquinamento sia molto sentito, la popolazione
attende soluzioni calate dall'alto, ma pu\`o comunque essere educata al
rispetto dell'ambiente, anzitutto tramite la conoscenza del territorio e
della sua natura, tra i quali spesso vive distrattamente. La cittadinanza
pu\`o non solo essere abituata al rispetto dell'ambiente, al riciclaggio dei
rifiuti, al riutilizzo dei materiali usati, al risparmio energetico, ma
pu\`o anche diventare un valido supporto alle Istituzioni nella lotta
all'inquinamento e al degrado del territorio che abita, se opportunamente
stimolata a difenderlo.

\section{L'idea.}

Simili premesse spingono le Associazioni {\it Galea} (Navigare nella
Cultura) e {\it Set} (Scienza, Educazione e Territorio) a muoversi al fine
di diffondere la cultura scientifica, coerentemente al loro operare in
quell'ambito con una particolare attenzione alla divulgazione. Le sue
caratteristiche, del resto, fanno di questa Citt\`a un contesto veramente
appetibile per chi voglia lavorare nel campo delle scienze naturali, fisiche
e matematiche, a vari livelli.

In aggiunta a tutto ci\`o dobbiamo tenere presente che Sesto Fiorentino si
prepara ad essere interessata da potenti novit\`a, che avranno a che
rivedere col suo ambiente e la fruizione della cultura scientifica.

\begin{itemize}
\item  Di qui a pochi anni si verificheranno delle trasformazioni ambientali
importanti, per l'avvio della costituzione del Parco della Piana e per il
recupero del Parco di Doccia, con conseguenti mutamenti della viabilit\`a.

\item  \`E ormai vicinissimo il trasferimento del Polo Scientifico nella
Piana, che porter\`a a Sesto Fiorentino il baricentro della cultura e della
ricerca scientifica universitaria in Toscana, oltre ad incrementarne
improvvisamente la popolazione studentesca.

\item  In seguito alla riforma dell'autonomia universitaria la vita e la
gestione dell'Universit\`a saranno fortemente modificate, le Facolt\`a
verranno forzate ad aprirsi a quelle parti del mondo civile ed economico in
grado di offrir loro servizi, collaborazioni, sinergie.

\item  In seguito alla riforma dell'autonomia scolastica assisteremo a
profondi cambiamenti nella vita e nella gestione delle scuole, che
similmente potranno costruire collaborazioni con tutti gli operatori
culturali esterni.
\end{itemize}

La comunit\`a sestese, la sua Amministrazione Comunale, pi\`u in generale le
sue Istituzioni e le sue realt\`a socioeconomiche, non dovrebbero essere
spettatori ma protagonisti di tutto ci\`o: protagonisti di una vera sfida
per far propria quella miniera di cultura, e di lavoro, che questi
cambiamenti potranno diventare!

Grande, si sa, \`e la potenzialit\`a del tessuto sociale sestese, lo
dimostra la sua storia; grande ha saputo essere la lungimiranza delle
associazioni, dell'Amministrazione Comunale e di molti operatori economici.
Importante \`e la presenza della Pubblica Istruzione, con un Liceo
Scientifico, un Liceo Artistico e un Istituto per Ragionieri. Infine,
grandissima \`e la generosit\`a di tutti quelli che, finora, nella cultura
hanno operato.

Forse per\`o per raccogliere quella sfida con successo occorre qualcosa di
ancora pi\`u organizzato, mirato e potenziato. Forse quel che oggi esiste
ancora non basta. Infatti:

\begin{itemize}
\item  non esiste un solo operatore culturale scientifico di dimensioni
importanti con il quale l'Amministrazione, le aziende, le scuole e
l'Universit\`a, dialoghino ed interagiscano;

\item  la conoscenza che ha il cittadino sestese medio dell'ambiente
naturale che lo circonda \`e comunque scarsissima, tantopi\`u in un periodo
di ''riflusso'' della cultura e della presenza ambientalista;

\item  le scuole non hanno ancora strumenti, programmi n\'e mentalit\`a tali
da integrarsi fruttuosamente con l'ambiente esterno: devono essere
''cercate'', ''stanate'' se occorre;

\item  il trasferimento delle facolt\`a scientifiche a Sesto \`e guardato
con sufficienza da parte dell'ambiente universitario, che non conosce la
realt\`a cittadina n\'e tantomeno ne \`e attratto. Del resto neanche la
realt\`a locale, Istituzioni e societ\`a civile, ha una piena coscienza
dell'opportunit\`a che l'incontro con il mondo accademico rappresenta.
\end{itemize}

Temiamo che se tale situazione non verr\`a rapidamente modificata resteranno
grandi opportunit\`a perdute l'incontro con le facolt\`a scientifiche,
cos\`\i\ come la realizzazione del Parco della Piana o la risistemazione di
quello della Collina. \`E quindi pienamente necessario lavorare attorno alla
problematica dell'offerta culturale in tema ambientale e scientifico.

\section{Descrizione del progetto.}

L'idea attorno alla quale stiamo lavorando, l'argomento di questo Progetto,
\`e la costituzione sul territorio del Comune di Sesto Fiorentino, di un
Centro per la Divulgazione della Cultura Scientifica (di seguito indicato
come Cdcs), ente autonomo ed autogestito, senza finalit\`a di lucro, che
cresca come patrimonio della Citt\`a sia per attivit\`a amatoriali (mostre,
cicli di conferenze, percorsi formativi, ricreazione scientifica), sia come
vera e propria azienda del privato sociale tesa a fornire servizi specifici
ad enti e cittadini.

Il Cdcs dovr\`a essere una struttura polifunzionale comprendente:

\begin{itemize}
\item  un laboratorio scientifico-naturalistico;

\item  una stazione per la produzione e la diffusione di materiali
multimediali;

\item  un ambiente in cui tenere seminari, proiettare diapositive e
audiovisivi;

\item  un punto di partenza per escursioni didattiche sul territorio;

\item  una mediateca permanente in cui materiale scientifico di varia natura
\`e custodito e consultabile dalla popolazione.
\end{itemize}

Naturalmente per tutto ci\`o \`e necessaria un'unica struttura. Accanto alla
sede del laboratorio scientifico-naturalistico (che per il suo peculiare
carattere necessita di un ambiente appositamente attrezzato) contiamo di
avvalerci della collaborazione di strutture gi\`a esistenti e gi\`a operanti
sul territorio come attrattori delle iniziative culturali. Segnatamente,
della Biblioteca Comunale, aperta nei mesi invernali anche nelle ore serali,
e perci\`o particolarmente adatta come punto di incontro per le riunioni
anche con adulti.

Descriviamo di seguito gli ambienti e le funzioni che richiediamo di poter
realizzare nella sede unica destinata al Cdcs dall'Amministrazione Comunale
di Sesto Fiorentino.

\subsection{Laboratorio scientifico ad indirizzo chimico, fisico e
matematico.}

Nell'ambito del progetto del Cdcs di Sesto Fiorentino riteniamo opportuna la
creazione di una laboratorio dedicato ad attivit\`a di educazione e
sperimentazione chimica, fisica e matematica rivolto sia alle scuole, sia al
singolo cittadino che abbia interesse nelle materie in questione.

Le funzioni che saranno accese all'interno di questa struttura saranno le
seguenti:

\begin{itemize}
\item  Esperimenti e dimostrazioni pratiche di facile comprensione di
carattere fisico, chimico, matematico eseguiti con strumenti scientifici da
personale qualificato.\\Tali esperimenti, a causa della complessit\`a di
realizzazione e dell'utilizzo di materiale non adatto in particolare agli
utenti pi\`u piccoli, sar\`a eseguito da operatori che illustreranno tutte
le fasi degli esperimenti in modo chiaro ed esauriente, eventualmente con
l'ausilio di supporti audiovisivi, dispense cartacee di nostra produzione ed
ipertesti come descritto in dettaglio successivamente.

\item  Esperimenti eseguibili dagli utenti stessi, da realizzarsi con
materiali di facile reperibilit\`a e di uso quotidiano, sotto la
supervisione del personale di laboratorio.\\Queste esperienze sono molto
utili per avvicinare il pubblico alle materie scientifiche passando per
esperienze di tutti i giorni in occasione delle quali entriamo costantemente
in contatto con la natura e le sue leggi. Le attivit\`a ora descritte
prevedono due fasi: la prima in cui il personale spiegher\`a sia i
fondamenti teorici alla base del fenomeno osservato, sia la realizzazione
dell'esperimento; e la seconda che consiste nella realizzazione pratica di
ci\`o che \`e stato appena appreso.\\Queste attivit\`a sono particolarmente
indicate per le scuole elementari, in quanto la seconda fase pu\`o
rappresentare un vero momento ludico d'apprendimento.

\item  Realizzazione di un centro di documentazione permanente sia
sull'attivit\`a svolta dal laboratorio stesso e dalle altre strutture
interne al Centro, sia delle iniziative a carattere scientifico che verranno
svolte sul territorio anche da altri Enti e strutture analoghe alla nostra,
in particolar modo dalle scuole elementari e medie inferiori.

\item  Realizzazione di un centro informazioni, con il compito di essere un
punto di riferimento per chiunque voglia svolgere attivit\`a scientifiche,
fornendo assistenza da parte di personale qualificato e strutture idonee
alla realizzazione di tali attivit\`a. In particolare tale centro
informazioni potr\`a essere uno strumento utile agli insegnanti che vogliono
interagire maggiormente con il Centro stesso.

\item  Punto d'incontro con le realt\`a a carattere scientifico, delle zone
limitrofe, per organizzare visite guidate a musei, laboratori, industrie,
centri di ricerca, osservatori astronomici ecc.

\item  Realizzazione di un laboratorio di matematica che si prefigga lo
scopo di avvicinare i ragazzi allo studio della matematica tramite
esperienze tattili e visuali, cercando di trasformare in gioco le
applicazioni matematiche. \`E nostro intento costruire strumenti didattici
in questo senso, che possano sia essere usati all'interno del Cdcs che dati
in prestito a scuole o a privati.
\end{itemize}

Il laboratorio vuole essere una realt\`a il pi\`u possibile aperta alla
cittadinanza, per questo motivo ricercher\`a la collaborazione di tutti
coloro che si dimostreranno interessati alle attivit\`a di cui sopra e
organizzer\`a anche iniziative al di fuori della propria sede in modo tale
da avvicinarsi maggiormente al singolo cittadino.

\subsection{Laboratorio scientifico ad indirizzo naturalistico, biologico ed
ecologico}

Il laboratorio, con caratteristiche polifunzionali e possibilit\`a di
fruizione costante, si propone di mettere a disposizione dell'utenza
attrezzature, materiali e personale qualificato allo scopo di:

\begin{itemize}
\item  Svolgere attivit\`a didattiche con scuole di ogni ordine e grado al
fine di acquisire, tramite esperienze di tipo naturalistico, nozioni
relative sia a metodologie agricole di tipo ecologico, sia volte alla
conoscenza dell'ambiente naturale del comprensorio;

\item  Creare laboratori per ragazzi e adulti dove realizzare, nel tempo
libero, esperienze di orticoltura biologica, giardinaggio, compostaggio,
riconoscimento di specie vegetali, fisiologia vegetale, erboristeria;

\item  Realizzare un centro di informazione e documentazione (dotato di
audiovisivi, libri, riviste specializzate, materiale multimediale...) volto
all'acquisizione di conoscenze di carattere storico e ambientale;

\item  Realizzare un centro di divulgazione di tecniche bioedilizie,
bioclimatiche, lagunaggio e risparmio delle risorse energetiche utilizzando
energie alternative e rinnovabili (solare, biogas etc.) per le quali esempio
concreto potr\`a essere la struttura stessa;

\item  Svolgere attivit\`a didattiche ed esplicative sui popolamenti della
''Piana'' e di Monte Morello, con particolare riferimento all'evidenze
archeologiche dello sviluppo demografico sin dalla preistoria.
\end{itemize}

Le attivit\`a da intraprendere nel laboratorio avranno carattere
interdisciplinare, per cui saranno condotte da esperti di ogni settore
naturalistico, (biologi, naturalisti, forestali, agronomi, ecc.) al fine di
fornire le pi\`u esaurienti spiegazioni in ogni campo
scientifico-naturalistico, e nel tentativo di dimostrare l'importanza e
l'utilit\`a di far comunicare tra loro pi\`u settori scientifici, sia
durante la ricerca, sia durante la divulgazione.

L'impostazione interdisciplinare di questa attivit\`a mira a creare, quindi,
una conoscenza esaustiva del mondo della natura perch\'e ragionata e
sviscerata sotto tutti gli aspetti.

A caratterizzare l'attivit\`a di educazione scientifica-naturalistica e di
conoscenza del territorio vi \`e una metodologia di base, che si fonda
sull'approccio diretto tra l'utente e l'argomento affrontato, e
sull'approfondimento differenziato per ogni categoria di destinatari:
scolari, studenti, anziani, o appassionati di ogni et\`a.

Nello svolgere ogni attivit\`a didattica ricreativa sar\`a sempre tenuto
presente lo scopo educativo, mirando alla comprensione degli effetti che
possono avere sugli ecosistemi determinate azioni dell'uomo.

Il centro fornisce il punto d'appoggio e la base per:

\begin{itemize}
\item  Realizzazione di attivit\`a orticole con le scuole materne,
elementari , medie e superiori

\item  Attivit\`a didattico-naturalistiche per scuole medie e superiori

\item  Ecoludoteca: attivit\`a ludiche per ragazzi nel tempo libero

\item  Attivit\`a pratiche per adolescenti, adulti, anziani anche a scopo
sociale ( categorie a rischio, disabili)

\item  Itinerari di scoperta delle caratteristiche storiche del territorio
della piana

\item  Percorsi naturalistici nelle zone umide della piana e lungo i canali

\item  Reperimento di informazioni riguardanti: vivaismo e agricoltura
biologica, compostaggio, bioarchitettura, educazione ambientale, aspetti
naturalistici del comprensorio

\item  Creazione di una rete di comunicazione con altri Centri di Educazione
Ambientale nel territorio nazionale (LABNET) e internazionale (INTERNET).
\end{itemize}

\subsection{Articolazione delle attivit\`a didattiche.}

Si programmano attivit\`a di divulgazione delle Scienze Naturali (qui di
seguito indicate con SN) rivolte sia alla cittadinanza in senso lato, sia a
settori di essa definiti da caratteristiche lavorative e sociali precise.
Queste attivit\`a sono riunite nei quattro settori Divulgativo, Scolastico,
Libera Universit\`a e Individuale di seguito illustrati, distinti dal bacino
di utenza al quale si rivolgono, e conseguentemente caratterizzati.

\subsubsection{Settore divulgativo.}

Il primo tipo di attivit\`a mira alla diffusione della conoscenza
scientifica delle SN a livello di divulgazione e si rivolge alla
cittadinanza indistintamente; sono attivit\`a pensate per fare delle SN una
occasione di socializzazione, di ricreazione e di approfondimento, e
destinate essenzialmente al tempo libero. L'obbiettivo \`e quello di far
entrare l'utente in contatto con le SN, con i problemi posti a livello
teorico, con lo stesso mondo di chi nelle SN opera professionalmente.

Le attivit\`a del settore divulgativo perci\`o non esigono una
programmazione organica complessiva, e vengono pensate e realizzate come
iniziative monografiche non esigenti una la frequenza dell'altra.

I contenuti scientifici con cui l'utente viene a contatto sono di
''difficolt\`a'' diversa, e non necessariamente piccola: non vi \`e alcuna
intenzione di relegare al settore divulgativo concetti, problemi e argomenti
necessariamente ''semplici''. Piuttosto \`e nella forma della loro
presentazione che si dovr\`a tener conto di come tali contenuti saranno
occasione di socializzazione e ricreazione ''culturale'' piuttosto che di
arricchimento di un bagaglio di interesse ''professionale''.

In tale ordine di idee si privileger\`a una presentazione discorsiva,
illustrativa, volta alla descrizione della fenomenologia pi\`u che alla
formalizzazione tecnica, tesa a far entrare l'utente in contatto con
l'accadere di un processo o col funzionare di un meccanismo. Il fenomeno
quindi \`e spiegato, in maniera anche serrata e soprattutto rigorosa, senza
concedere alcunch\'e alla spettacolarizzazione gratuita, ma facendo
necessariamente a meno di formalizzazione tecnica e matematica che interessa
soltanto un approccio professionale.

Le attivit\`a divulgative del Cdcs possono essere riunite in quattro tipi a
seconda del luogo fisico dove si svolgono e del personale a cura di cui sono.

\paragraph{Incontri monografici divulgativi.}

Si tengono lezioni monografiche di SN, incentrate soprattutto su immagini da
proiettarsi con lavagna luminosa e proiettore, si visiona e discute
materiale in videocassetta, si eseguono e spiegano dimostrazioni ed
esperimenti tratti dal programma per le scuole dell'obbligo che siano
ritenuti particolarmente significativi per i non addetti.

Si propongono anche attivit\`a divulgative che s'avvalgano dell'apporto di
esterni alle Associazioni titolari del progetto in qualit\`a di animatori
scientifici. Si pensa a ricercatori, scienziati, tecnici e pubblicisti in
genere invitati dal Cdcs a presentare, nel contesto e nello spirito delle
Attivit\`a divulgative, propri lavori, risultati, pubblicazioni ritenuti di
interesse.

Questi incontri si programmano sia nella stessa sede del Centro, che in
altre sedi di pubblico interesse, a cura di animatori scientifici interni
alle Associazioni titolari del progetto Cdcs od in collaborazione con
personale di dette altre sedi. Segnatamente si rivolge questa proposta di
attivit\`a divulgativa a: biblioteche, circoli aderenti all'Associazionismo
Sestese, centri sociali, circoli parrocchiali, circoli culturali e sportivi
privati, aziende, locali, centri commerciali, Asl, circoli dopolavoro ed
altri di simile carattere e interesse.

\paragraph{Gite d'istruzione.}

Collaborando con gli operatori scientifici della divulgazione direttamente
collegati con l'universit\`a ed i centri di ricerca, si effettuano visite
guidate ad osservatori astrofisici, laboratori nazionali ed europei, musei,
parchi naturali, acquari, od escursioni in luoghi panoramici con
osservazioni astronomici.

I responsabili scientifici delle Gite di istruzione, che potranno essere
interni o meno alle Associazioni titolari del progetto, saranno designati
secondo il Regolamento del Cdcs.

\paragraph{Produzione di materiale didattico.}

Il contenuto delle diverse iniziative scientifiche didattiche sar\`a anche
presentato in forma di dispense monografiche autoprodotte dai soci delle
Associazioni titolari del progetto, che potranno essere in queste forme
sottoelencate od in altre forme.

\begin{itemize}
\item  Dispense monografiche cartacee.

\item  Dispense interattive in forma di ipertesto su cd rom o floppy disk.

\item  Collezioni di diapositive.

\item  Collezioni di fotografie.

\item  Videodispense.

\item  Audiodispense.
\end{itemize}

Detto materiale divulgativo sar\`a commercializzato secondo la legge sulle
Societ\`a Non-Profit ed il ricavato verr\`a reinvestito nella gestione del
Cdcs secondo il Regolamento e le deliberazioni del Consiglio
d'Amministrazione del Centro.

\subsubsection{Attivit\`a scolastiche.}

Capitolo fondamentale della programmazione di SN \`e naturalmente il lavoro
negli Istituti scolastici, anche in previsione dell'attuazione della riforma
della loro autonomia. Diverse tipologie di attivit\`a si possono individuare:

\begin{itemize}
\item  cicli di lezioni di Fisica moderna (Meccanica quantistica e
Relativit\`a) che portano nella scuola argomenti quasi totalmente estranei
dagli attuali programmi ministeriali;

\item  lezioni di assistenza alla programmazione istituzionale: potremmo
offrire agli insegnanti l'occasione di alleggerire la propria programmazione
curando approfondimenti e ripassi su argomenti in programma;

\item  corsi speciali per gruppi di studenti che desiderino approfondire
particolarmente la materia;

\item  esperimenti, all'interno del laboratorio del Cdcs, pensati e
realizzati con gli insegnanti in modo da rispondere a precise esigenze di
programma; (seminari sugli ecosistemi forestali e le loro specie in forma
monografica e/o consequenziale (tipo corso) mirati ad approfondimenti
particolari; (corsi di ecologia, botanica, zoologia, monografici e/o
consequenziale (tipo corso) mirati pi\`u che alla formazione scientifica
professionale alla presa di coscienza, da parte dei corsisti, della
complessit\`a e delicatezza degli equilibri degli ecosistemi (gite e
trekking naturalistici allo scopo di osservare gli ecosistemi, le loro
componenti e le loro dinamiche.
\end{itemize}

\subsubsection{Libera Universit\`a.}

Nei locali del Cdcs si organizzano corsi autogestiti di programma fissato
dai titolari designati secondo il Regolamento del Centro fra soci interni o
personale esterno; questi corsi costituiscono nel loro complesso un ciclo
curricolare vero e proprio a cui diamo il nome di Libera Universit\`a.

Si tratta dell'ambizioso e sentito progetto di creare, esternamente al ciclo
istituzionale degli studi scolastici ed universitari, l'occasione per lo
studio di livello professionale o semi-professionale delle SN, da offrire ai
cittadini che, per vicende diverse (condizioni economiche, situazioni
storiche, scelta lavorativa, altre esigenze) non hanno potuto frequentare i
corsi universitari pur avendo interesse ed inclinazione alla materia.

La Libera Universit\`a prevede l'iscrizione e la frequenza a cicli di
lezioni, dalla quale ci si aspetta non solo l'assimilazione di nozioni e
concetti inerenti alla disciplina studiata, ma anche l'appropriazione di un
certo bagaglio tecnico potenzialmente utilizzabile in sede professionale.

\subsubsection{Attivit\`a individuali.}

Le attivit\`a individuali sono percorsi culturali rivolti al cittadino come
lezioni individuali o momenti autodidattici, in cui egli \`e seguito dal
nostro personale qualificato. In esse si distinguono Doposcuola scientifico
(rivolto a studenti ed alunni della scuola pubblica e privata che vogliano
supportare od approfondire la loro preparazione), Percorsi professionali
(studiati apposta per le esigenze specifiche di lavoratori) e Ludoteca
scientifica (rivolta a bambini, ragazzi e adulti che vogliano avvicinarsi,
attraverso esperimenti e audiovisivi, al mondo delle SN).

\paragraph{Doposcuola scientifico.}

L'offerta scolastica pubblica e privata nel nostro Paese ha grandi meriti e
limiti sui quali sarebbe qui molto lungo distendersi, e che non \`e nostra
intenzione criticare.

\`E tuttavia un dato che un bacino d'utenza vastissimo fra gli studenti e
gli alunni delle scuole italiane ricorre, per motivi diversissimi e con un
panorama di investimenti quantomai variegato, a lezioni private di recupero
e supporto. \`E un altro dato inconfutabile che l'offerta di lezioni private
resta per la quasi totalit\`a al di fuori di ogni legislazione, ed \`e per
questo lasciata quasi completamente all'arbitrio di una contrattazione
insegnante-studente che funziona come un doppio ricatto: chi offre lezioni
private lo fa prevalentemente per mancanza di altri redditi regolari, che le
richiede lo fa per l'incapacit\`a, un po' sua un po' della struttura
scolastica, di far s\`\i\ che egli apprenda le nozioni necessarie al suo
curriculum per via ''istituzionale''.

Cos\`\i\ chi ha bisogno di questo tipo di supporto didattico deve affidarsi
o ad un mercato clandestino dai prezzi completamente fluttuanti (e dal
personale docente la cui preparazione \`e in generale fuori dal controllo
dell'utente), o ad istituti privati che operano s\`\i\ alla luce del sole,
ma con tariffe spesso violentemente proibitive. Inutile sottolineare che
questa situazione rappresenta ipso facto una inaccettabile selezione
censitaria degli studenti!

L'intervento pubblico che mette in moto la Scuola di Stato pu\`o tentare di
non arrestarsi fuori dagli Istituti, ed escogitare, per mezzo dell'Ente
Locale, forme di aiuto agli studenti in difficolt\`a, offrendo loro una
valida alternativa al mercato esoso delle lezioni private. Naturalmente
ci\`o deve avvenire in forma dinamica e non assistenziale: nel caso del
progetto che proponiamo l'Amministrazione Comunale promuove il Cdcs che
garantisce un servizio ''municipalizzato'' nelle materie scientifiche di
lezioni individuali, lavorando con tariffe studiate per rendere accessibile
al massimo numero di utenti un servizio dai costi attualmente gravosi.

Pensiamo non tanto di sostituire alla lezione privata la lezione
''municipalizzata'' fatta a domicilio del singolo studente: piuttosto
realizzeremo un ''doposcuola'' in cui lo studente possa essere seguito
singolarmente, con la stessa attenzione che riceverebbe nella lezione
privata, ma potendo usufruire della sede del Cdcs, dei materiali didattici e
scolastici ivi utilizzabili! Studiando le caratteristiche e le esigenze
della preparazione degli studenti in difficolt\`a che al Cdcs si rivolgano,
gli insegnanti del Centro incoraggeranno gli stessi a discutere fra di loro
delle materie esaminate, formando gruppi in cui allo studio individuale e
rigoroso seguano confronti e dibattiti sugli argomenti approfonditi.

Gli insegnanti verranno designati attraverso voto del Comitato di Gestione
del Cdcs conformemente al Regolamento del Centro, fra personale iscritto
alle Associazioni titolari del progetto o cittadini non iscritti e
lavoreranno come liberi professionisti, nei termini della legge.

\paragraph{Percorsi professionali individuali.}

Per lavoratori e/o disoccupati che abbiano interesse di migliorare le loro
conoscenze in campo scientifico naturalistico, al fine di qualificare le
loro prestazioni professionali sia autonome che dipendenti. Tali percorsi
avranno sempre carattere pratico-applicativo in modo che l'utente possa
attuare subito le conoscenze acquisite.

\paragraph{Ludoteca scientifico-naturalistica.}

La ludoteca da ospitare negli stessi locali del laboratorio ma in orari
diversi, si rivolge a tutta la popolazione, dando la possibilit\`a a
chiunque ne faccia richiesta di poter eseguire o assistere agli esperimenti
desiderati, di poter consultare tutti i materiali multimediali presenti, di
ricevere delucidazioni in materia scientifica da parte degli operatori
presenti. Questo, oltre a costituire un momento di gioco per i ragazzi,
potr\`a essere un'occasione per coloro che gi\`a si interessano di scienza a
livello non professionale per approfondire le proprie conoscenze, e per
coloro che ne sono completamente digiuni per cominciare ad interessarsene.

Vuole essere, inoltre, il luogo dove i pi\`u piccoli abbiano la
possibilit\`a di iniziare a conoscere la natura in modo diretto attraverso
esperienze sensoriali, ed indiretto con giochi, immagini e audiovisivi
aventi per protagonisti soggetti presenti in natura.

La caratteristica di tali attivit\`a sar\`a quella di fare entrare in
contatto i piccoli allievi con il mondo della natura attraverso i cinque
sensi educandoli attraverso il gioco, alla conoscenza e alla curiosit\`a
verso il mondo che li circonda e soprattutto al rispetto per l'ambiente in
senso lato.

\section{Risultati attesi.}

Il centro di didattica ambientale contribuir\`a direttamente ed
indirettamente con le sue funzioni ad un incremento occupazionale, anche se
sono previste prestazioni volontaristiche e contributi pubblici, provvedendo
ad animare corsi di formazione, di aggiornamento, di divulgazione, di
ricerca, di informazione (anche attraverso le reti telematiche e i network a
carattere ecologico).

\section{Carattere innovativo.}

I mezzi impiegati consentiranno di caratterizzare in modo nuovo
l'impostazione della nostra attivit\`a, che vuole essere tale da non
trascurare e tralasciare nessun tipo di utente, e approfondire
esaurientemente gli argomenti oggetto di studio in modo appropriato alle
esigenze dei partecipanti. Questo implica non solo la produzione e
diffusione, attraverso pubblicazioni di vario genere (botanico, zoologico,
paesaggistico, turistico-naturalistico, materiali ricreativi e didattici) di
nozioni ecologiche e scientifico-naturalistiche, ma anche la volont\`a di
far crescere nelle persone una nuova coscienza e conoscenza ecologica.

Il Cdcs \`e nuovo anche per la mentalit\`a o forse meglio per
l'atteggiamento verso il mondo delle scienze, infatti mira a riunire le
varie discipline scientifiche, in particolare quelle relative allo studio
dei fenomeni naturali sia, durante la ricerca sia, durante la discussione
sia, durante la divulgazione.

In ultimo la grande novit\`a sta nell'obbiettivo, tanto arduo quanto
importante ed innovativo, di guardare all'educazione ambientale e
all'educazione alla scoperta del territorio, come al mezzo per formare una
conoscenza ed una coscienza ecologica nei cittadini, da tradurre poi in uno
spontaneo stile di vita ecocompatibile da parte di tutta la comunit\`a.

\section{Riproducibilit\`a del progetto.}

Il centro didattico, oltre ad offrire un supporto culturale e scientifico,
\`e al contempo un luogo di incontro e di scambio sociale, per cui, questo
tipo di funzione, \`e estendibile a centri analoghi rivolti ad altri settori
di interesse, rispetto al patrimonio storico-culturale, architettonico,
archeologico, ecc., funzionando come poli di rivitalizzazione creativa sul
territorio e nelle interrelazioni e negli scambi sociali.

La previsione di un piano variamente articolato ed organico nel suo insieme,
assicurano un giusto inserimento nel territorio anche ad interventi
specialistici come oasi faunistiche, parchi naturali, riserve di caccia,.
impianti di lagunaggio, ecc., permettendo di soddisfare funzioni e bisogni
diversi, stemperando tensioni e diversi bisogni che attualmente si
contendono le poche aree esistenti disponibili.

\section{Conclusioni.}

Le Associazioni titolari di questo progetto non nascondono di aver in mente
di costruire pezzo su pezzo una vera e propria occasione di lavoro per
giovani studiosi che, in futuro, possano farne un'attivit\`a professionale a
tempo pieno.

Conformemente alle ispirazioni ed alle finalit\`a sociali previste dai loro
statuti, le Associazioni Galea e Set puntano su questo progetto per
realizzare, nel Comune di Sesto Fiorentino, insieme una struttura che
divenga prezioso patrimonio per la collettivit\`a cittadina e ed occasione
per creare lavoro nei ''servizi sociali'' dell'ambito culturale. Cos\`\i\
quello che oggi \`e un contributo per attivit\`a culturali-ricreative nel
settore propriamente no - profit, l'avvio per un'attivit\`a associativa di
un gruppo di appassionati, potrebbe trasformarsi, crescendo, in un
lungimirante investimento capace di fruttificare in termini di qualit\`a
della vita a Sesto, fruibilit\`a della cultura ed occupazione.

Guardiamo alla costruzione del Cdcs con grande entusiasmo, anzitutto per un
fatto sociale: incominciare a parlarne, a stendere il presente progetto,
cercare di disegnarlo nelle nostre aspirazioni, nella nostra immaginazione,
confrontandolo con le situazioni oggettive, con gli insegnamenti che ognuno
di noi trae dalle proprie esperienze passate, \`e stata un'operazione di
grande valore umano ed organizzativo. Per la prima volta per quel che ne
sappiamo, una Associazione preesistente (Galea) ed una di nuova formazione
(Set), si sono messe attorno a un tavolo per sperimentare una creazione
comune, un progetto unificante, invertendo il processo di frammentazione
latente che spesso arricchisce il panorama associativo disgregando realt\`a
esistenti. Ci \`e parsa una strada da percorrere, anche contro l'orgoglio
individuale delle singole associazioni che partecipano, anche contro la
gelosia della propria identit\`a, puntando pi\`u sulla costruzione comune,
sul servizio che solo uniti si pu\`o fornire, che su quanto bravi si pu\`o
essere ognuno per s\'e.

Ci \`e parso un progetto ambizioso e lontano, e proprio per questo da
perseguire!

E sicuramente ancora pi\`u ambizioso \`e il progetto di far crescere il Cdcs
al punto di diventare, in un futuro lontano che lavoreremo per avvicinare,
una vera occasione lavorativa, per impiegarsi in un ambito che ci \`e
congeniale, rendendoci utili al territorio, e padroni del nostro lavoro.

\end{document}